\documentclass{PoS}
\usepackage[italian,english]{babel}
\usepackage{amssymb}
\usepackage{amsfonts}
\usepackage{amsmath}
\usepackage{fancyhdr}
\usepackage{lineno}
\usepackage{subfigure}
\newcommand{\nn}{\nonumber}
\newcommand{\lsim}{\mathrel{\mathop{\kern 0pt \rlap
  {\raise.2ex\hbox{$<$}}}
  \lower.9ex\hbox{\kern-.190em $\sim$}}}
\newcommand{\gsim}{\mathrel{\mathop{\kern 0pt \rlap
  {\raise.2ex\hbox{$>$}}}
  \lower.9ex\hbox{\kern-.190em $\sim$}}}

\newcommand{\be}{\begin{equation}}
\newcommand{\ee}{\end{equation}}
\newcommand{\bea}{\begin{eqnarray}}
\newcommand{\eea}{\end{eqnarray}}

\usepackage{pstricks}
\usepackage{color}
\usepackage{multirow}
\usepackage{cite}

\title{Higgs bosons: discovered and hidden,  in extended Supersymmetric Standard Models at the LHC}

\ShortTitle{Higgs bosons: discovered and hidden Higgs Bosons in extended Supersymmetric Standard Models at the LHC}

\author{\speaker{Priyotosh Bandyopadhyay}%
\\
        INFN Lecce, 
        Dipartimento di Matematica e Fisica "Ennio De Giorgi", \\ Universit\`a del Salento and INFN-Lecce, \\ Via Arnesano, 73100 Lecce, Italy\\

        E-mail: \email{priyo@le.infn.it, bpriyo@gmail.com}}

\author{$^{(a,b)}$ Claudio Corian\`o and  $^{(a)}$Antonio Costantini\\

       $^{(a)}$INFN Lecce, 
        Dipartimento di Matematica e Fisica "Ennio De Giorgi", \\ Universit\`a del Salento and INFN-Lecce, \\ Via Arnesano, 73100 Lecce, Italy\\

$^{(b)}$STAG Research Centre and Mathematical Sciences,
University of Southampton, Southampton SO17 1BJ, UK\\
        E-mail: \email{claudio.coriano@le.infn.it, antonio.costantini@le.infn.it\\}}

\abstract{  We investigate an extension of the Minimal
Supersymmetric Standard Model (MSSM) containing a $SU(2)$ Higgs triplet of zero hypercharge and a gauge singlet.  We focus on a scenario of this model which allows a light pseudoscalar and/or a scalar
below $100$ GeV in the spectrum, consistent with the most recent data from the LHC and the earlier data from the LEP experiments. We analyze the exotic decay of the discovered Higgs $(h_{125})$ into two light (hidden) Higgs bosons present in the extension. The latter are allowed by the uncertainties in the Higgs decay  $h_{125}\to WW^*$, $h_{125}\to ZZ^*$ and $h_{125}\to \gamma\gamma$. We have searched for such light Higgs bosons in the $2b+2\tau$, $\geq 3\tau$, $2b+2\mu$ and $2\tau+2\mu$ final states at the LHC with 13 and 14 TeV. A region of such parameter space can be explored with an integrated luminosity of 25 fb$^{-1}$ at the LHC.}

\FullConference{Proceedings of the Corfu Summer Institute 2015 "School and Workshops on Elementary Particle Physics and Gravity"\\

                 1-27 September 2015\\

                 Corfu, Greece}

\begin{document}

\section{Introduction}
The discovery of the Higgs boson around 125 GeV has certainly proved the existence of at least one scalar taking part in the 
process of electroweak symmetry breaking (EWSB). The presence of such a scalar has been experimentally 
verified in its decay modes into $WW^*$,
$ZZ^*$ and $\gamma\gamma$ \cite{CMS} at the Large Hadron Collider (LHC).  Other Standard Model (SM) decay modes, i.e., to $b\bar{b}$, $\tau\bar{\tau}$ and $\mu\bar{\mu}$, have yet to reach the $5\sigma$ statistical significance. In the midst of searching for SM decay modes, it is also important to investigate other non standard searches.  Such non-standard decay modes are often motivated by various theoretical constructs. 
The SM is a very successful theory, but it has also some well-known shortcomings, such as the gauge hierarchy problem in the Higgs sector or the absence of a cold dark matter candidate in its spectrum. At the same time it falls short from offering a justification for the mass of the light neutrinos. Supersymmetry has been one of the most popular scenario which may provide 
an answer to several of these issues and defines a significant framework for addressing physics beyond the SM. One of its specific features is in the presence of one additional Higgs doublet, in its minimal formulation. \\
In this work we are going to illustrate the theoretical possibilities which remain, at this time, wide open from a phenomenological perspective and which define possible scenarios for an extended Higgs sector, in the context of supersymmetry. These constructions also address other theoretical issues, with exciting phenomenological implications. The extension of the Higgs sector with a singlet superfield can solve the
$\mu$ problem in the minimal supersymmetric SM (MSSM) dynamically.  Similarly, for the same reason, one can introduce an $SU(2)$ triplet superfield, but in this case such dynamical solution is not possible, due to the constraints imposed by the $\rho$ parameter on the vacuum expectation value (v.e.v.) of the triplet \cite{rho}. Extending the Higgs sector by a singlet/triplet also gives the possibility of a spontaneous violation of the CP symmetry.\\
Here, for simplicity, we consider a supersymmetric extension of the Higgs sector by a $SU(2)\times U(1)_Y$ singlet and a $Y=0$ hypercharge $SU(2)$ triplet. We will see
that the additional $Z_3$ symmetry of the Lagrangean may predict the existence of a light pseudoscalar. Such possibilities can lead to additional decay modes of the discovered Higgs boson around 125 GeV into a pair of light pseudoscalars and into a light pseudoscalar accompanied by a $Z$ boson, if these are kinematically allowed.  \\
The charged Higgs sector also gets affected due to the existence of a light pseudoscalar,  with a light charged Higgs that may now decay into a pseudoscalar and one $W^\pm$ boson. This particular decay is allowed for both doublet and triplet-like charged Higgs bosons. However, the $Y=0$ triplet brings in an additional coupling of the charged Higgs boson to the $Z$ and $W^\pm$ at tree-level, which breaks the custodial symmetry. This can generate a totally new decay mode for the charged Higgs boson into $Z\, W^\pm$, which is not present in the 2-Higgs doublet model (2HDM) or in the MSSM. Here we are going to discuss such possibilities and elaborate on how to explore them at current and future experiments. \\
In section~\ref{Model} we are going to briefly illustrate the model, moving to a discussion of its mass spectrum in section~\ref{1loop}. The perturbativity of the couplings is discussed in section~\ref{pblmt}. In section~\ref{lpss} we address the possibility of having a hidden scalar in the spectrum, while the phenomenological aspect of this scenario will be considered in section~\ref{pheno}. Our conclusions will be presented in section~\ref{concl}. 

\section{Model}\label{Model}
As explained in \cite{TNSSMo} the model  contains a SU(2) triplet $\hat{T}$ of zero hypercharge ($Y=0$)  together with a SM gauge singlet ${\hat S}$, added to the superfield content of the MSSM. The structure of its superpotential can be decomposed in the form 
 \begin{equation}
 W_{TNMSSM}=W_{MSSM} + W_{TS},
 \end{equation}
with
\begin{equation}
W_{MSSM}= y_t \hat U \hat H_u\!\cdot\! \hat Q - y_b \hat D \hat H_d\!\cdot\! \hat Q - y_\tau \hat E \hat H_d\!\cdot\! \hat L\ ,
\label{spm}
 \end{equation}
 being the Yukawa part of the MSSM superpotential, while 
 \begin{equation}
W_{TS}=\lambda_T  \hat H_d \cdot \hat T  \hat H_u\, + \, \lambda_S \hat S\,  \hat H_d \cdot  \hat H_u\,+ \frac{\kappa}{3}\hat S^3\,+\,\lambda_{TS} \hat S \, \textrm{Tr}[\hat T^2]
\label{spt}
 \end{equation}
accounts for the extended scalar sector, which includes an electroweak triplet and a singlet superfields. In our notation a ''$\cdot$'' denotes a contraction with the Levi-Civita symbol $\epsilon^{ij}$, with $\epsilon^{12}=+1$  The triplet and doublet superfields are given by 

\begin{equation}\label{spf}
 \hat T = \begin{pmatrix}
       \sqrt{\frac{1}{2}}\hat T^0 & \hat T_2^+ \cr
      \hat T_1^- & -\sqrt{\frac{1}{2}}\hat T^0
       \end{pmatrix},\qquad \hat{H}_u= \begin{pmatrix}
      \hat H_u^+  \cr
       \hat H^0_u
       \end{pmatrix},\qquad \hat{H}_d= \begin{pmatrix}
      \hat H_d^0  \cr
       \hat H^-_d
       \end{pmatrix}.
 \end{equation}
 Here $\hat T^0$ denotes a complex neutral superfield, while  $\hat T_1^-$ and $\hat T_2^+$ are the charged Higgs superfields.  
 The MSSM Higgs doublets are the only superfields which couple to the fermion multiplet via Yukawa coupling, as in Eq.~(\ref{spm}). The singlet and the triplet superfields account for the supersymmetric $\mu_D$ term, which couples $H_u$ and $H_d$ after that their neutral components acquire vacuum expectation values in Eq.~(\ref{spt}).
 
It is a characteristic of any scale invariant supersymmetric theory with a cubic superpotential that the complete Lagrangian with the soft SUSY breaking terms carries an accidental  $Z_3$ symmetry. This is generated by the invariance of all of its components 
after multiplication of the chiral superfields by the phase $e^{2\pi i/3}$ which, as we are going to discuss below, affects the mass of the pseudoscalars. 

The soft breaking terms in the scalar potential are given by

 \bea\nn
V_{soft}& =&m^2_{H_u}|H_u|^2\, +\, m^2_{H_d}|H_d|^2\, +\, m^2_{S}|S|^2\, +\, m^2_{T}|T|^2\,+\, m^2_{Q}|Q|^2 + m^2_{U}|U|^2\,+\,m^2_{D}|D|^2 \\ \nn
&&+(A_S S H_d\cdot H_u\, +\, A_{\kappa} S^3\, +\, A_T H_d\cdot T\cdot H_u \, +\, A_{TS} S\, Tr(T^2)\\ 
 &&\,+\, A_U U H_U \cdot Q\, +\, \, A_D D H_D \cdot Q + h.c),
\label{softp}
 \eea
while the D-terms take the form 
 \begin{equation}
 V_D=\frac{1}{2}\sum_k g^2_k ({ \phi^\dagger_i t^a_{ij} \phi_j} )^2 .
 \label{dterm}
 \end{equation}
All the coefficients involved in the Higgs sector are chosen to be real in order to preserve CP invariance. The breaking of the $SU(2)_L\times U(1)_Y$ electroweak symmetry is then obtained by giving real vevs to the neutral components of the Higgs field
 \be
 <H^0_u>=\frac{v_u}{\sqrt{2}}, \, \quad \, <H^0_d>=\frac{v_d}{\sqrt{2}}, \quad <S>=\frac{v_S}{\sqrt{2}}, \, \quad\, <T^0>=\frac{v_T}{\sqrt{2}},
 \ee
 which give mass to the $W^\pm$ and $Z$ bosons
 \be
 m^2_W=\frac{1}{4}g^2_L(v^2 + 4v^2_T), \, \quad\ m^2_Z=\frac{1}{4}(g^2_L \, +\, g^2_Y)v^2, \, \quad v^2=(v^2_u\, +\, v^2_d), 
\quad\tan\beta=\frac{v_u}{v_d} \ee
 and also induce, as mentioned above, a $\mu$-term of the form $ \mu_D=\frac{\lambda_S}{\sqrt 2} v_S+ \frac{\lambda_T}{2} v_T$.
 
The triplet vev $v_T$ is strongly  constrained by the global fit to the measured value of the $\rho$ parameter \cite{rho}
 \be
 \rho =1.0004^{+0.0003}_{-0.0004} ,
 \ee 
 which restricts its value to $v_T \leq 5$ GeV. Respect to the tree-level expression, the non-zero triplet contribution to the $W^\pm$ mass leads to a deviation of the $\rho$ parameter
 \be
 \rho= 1+ 4\frac{v^2_T}{v^2} .
 \ee
In the numerical analysis we have chosen $v_T =3$ GeV. This makes the effective value of $\mu_D$ coming from the triplet contribution rather low in order to satisfy the phenomenological constraints. On the other hand, the singlet vev $v_S$ is not affected by any such bound and can generate a value of $\mu_D$ which can be very large. However,  in gauged $U(1)'$ models where the singlet is responsible for the mass of the $Z'$, the singlet vev is subjected to an additional bound from the lower limit of the $Z'$ mass.

In the TNMSSM, the neutral CP-even mass matrix is $4$-by-$4$, since the mixing terms involve the two $SU(2)$ Higgs doublets, the scalar singlet $S$ and the neutral component of the Higgs triplet.  After electroweak symmetry breaking, the neutral Goldstone gives mass to the $Z$ boson, while the charged Goldstone bosons give mass to the $W^\pm$ boson. Being the Lagrangean CP-symmetric, we are left with four CP-even, three CP-odd  and three charged Higgs bosons as shown below
 \bea\label{hspc}
  \rm{CP-even} &&\quad \quad  \rm{CP-odd} \quad\quad   \rm{charged}\nn \\
 h_1, h_2, h_3, h_4 &&\quad \quad a_1, a_2, a_3\quad \quad h^\pm_1, h^\pm_2, h^\pm_3. 
 \eea
The neutral Higgs bosons are linear combinations of doublets, triplet and singlets, whereas the charged Higgses are  combinations of doublets and of a triplet only. We will denote with $m_{h_i}$ the corresponding mass eigenvalues, assuming that one of them will coincide with the 125 GeV Higgs $(h_{125})$ boson detected at the LHC. We investigate the scenario where one 
(or more) scalar or a pseudoscalar with a mass $< 125$ GeV is allowed, which we call a {\em hidden Higgs} scenario.

\begin{figure}
\begin{center}
\includegraphics[width=0.6\linewidth]{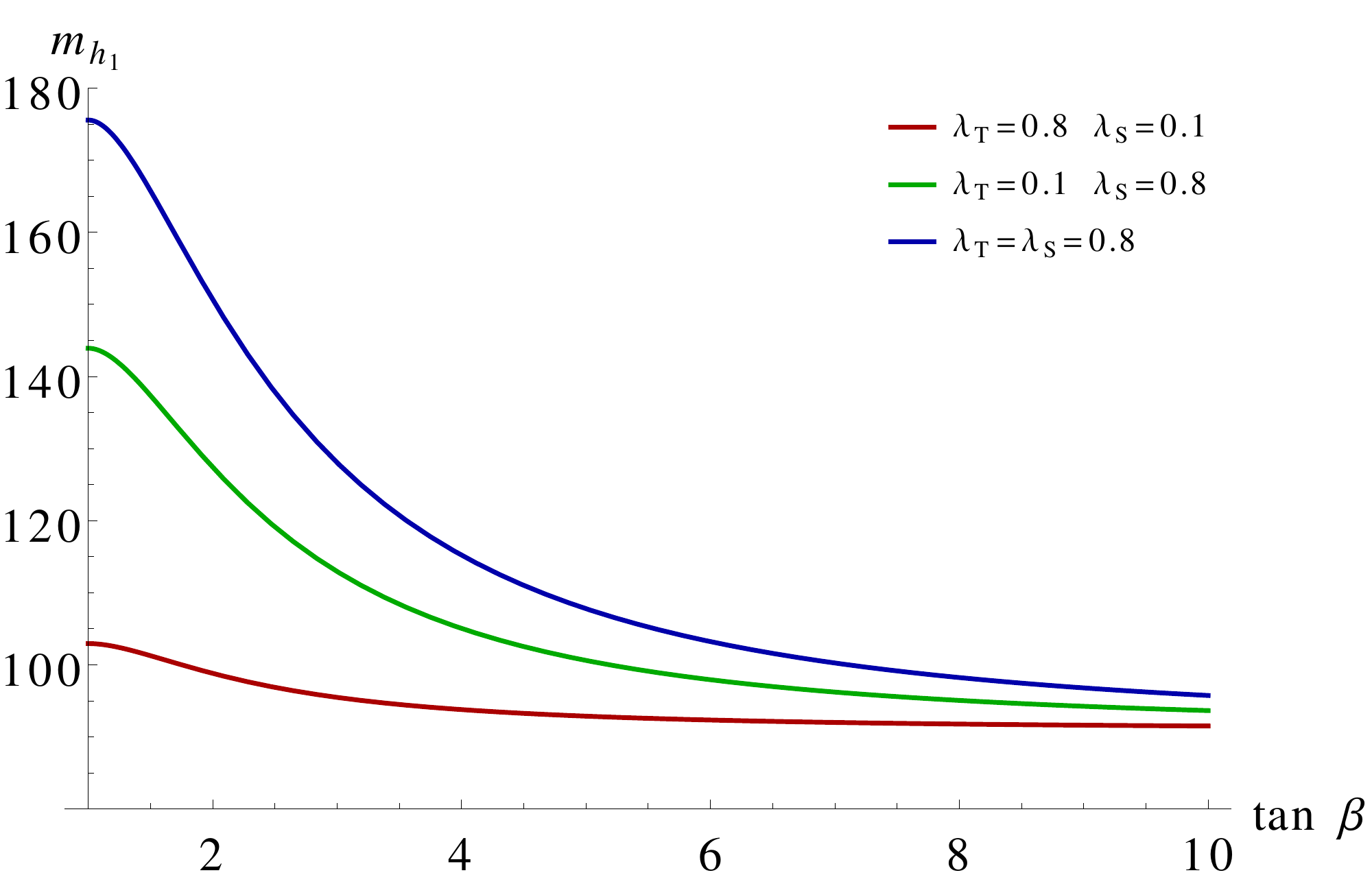}
\caption{Tree-level lightest CP-even Higgs mass maximum values versus $\tan{\beta}$ for 
 (i) $\lambda_T=0.8,\, \lambda_S=0.1$ (in red), (ii)$\lambda_T=0.1,\, \lambda_S=0.8$ (in green ) and (iii) $\lambda_T=0.8,\, \lambda_S=0.8$ (in blue) \cite{TNSSMo}.}\label{mht}
\end{center}
\end{figure}

At tree-level the maximum value of the lightest neutral Higgs has additional contributions from the triplet and the singlet sectors respectively. The numerical value of the upper bound on the lightest CP-even Higgs can be extracted from the relation
\be\label{hbnd}
m^2_{h_1}\leq m^2_Z(\cos^2{2\beta} \, +\, \frac{\lambda^2_T}{g^2_L\,+\,g^2_Y }\sin^2{2\beta}\, +\, \frac{2\lambda^2_S}{g^2_L\,+\,g^2_Y }\sin^2{2\beta}),
\ee
which is affected on its right-hand-side by two additional contributions from the triplet and the singlet. These can raise the allowed tree-level Higgs mass. Both contributions are proportional to $\sin{2\beta}$, and thus they can be large for a low value of $\tan{\beta}$, as shown in Figure~\ref{mht}. The additional contributions coming from the triplet and the singlet reduce the fine-tuning of the supersymmetric mass scale 
required to attain the lightest CP-even Higgs boson mass of 125 GeV. These extra contributions at tree-level are large for a low value of $\tan{\beta}$, and so we do not need radiative corrections in order to match the observed mass. However, 
for a large $\tan{\beta}$ value, these extra scalars contribute enough at higher orders, reducing the radiative corrections coming from the squarks and hence the required supersymmetric mass scale, which could be much below the TeV range.

\section{One loop Higgs boson masses}\label{1loop}

\begin{figure}
\begin{center}
\includegraphics[width=0.6\linewidth]{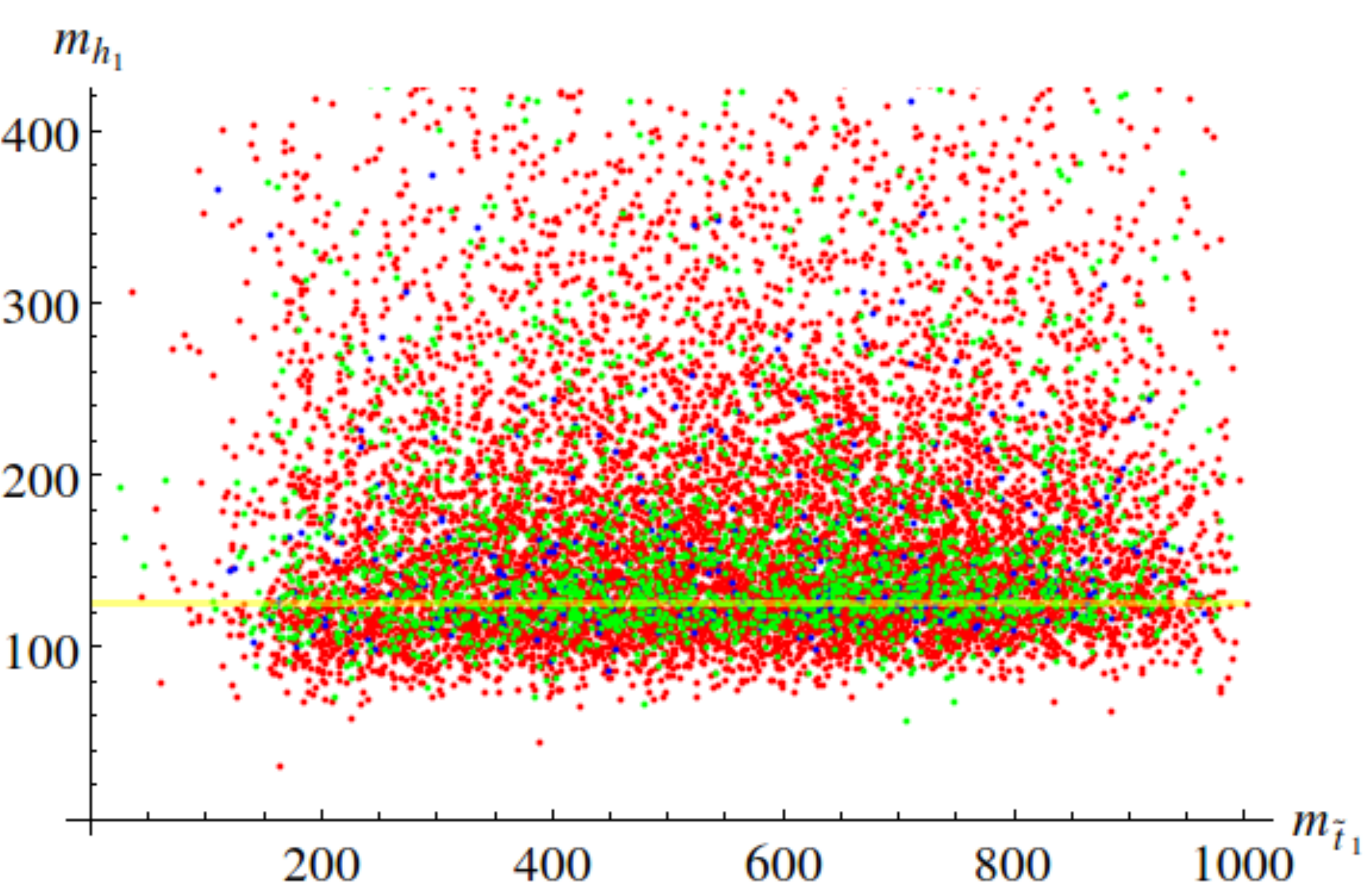}
\caption{The variation of  the one-loop lightest CP-even Higgs mass $m_{h_1}$ with  the lightest stop mass $m_{\tilde{t}_1}$. The yellow band shows the candidate Higgs mass $123\leq m_{h_1} \leq 127$ GeV \cite{TNSSMo}.}\label{mh1lvsp}
\end{center}
\end{figure}

To study the effect of the radiative correction to the Higgs masses, we calculate the one-loop
Higgs mass for the neutral Higgs bosons via the Coleman-Weinberg effective 
potential \cite{Coleman:1973jx} given in Eq.~(\ref{cwe}) 
\begin{align}\label{cwe}
V_{\rm CW}=\frac{1}{64\pi^2}{\rm STr}\left[ \mathcal{M}^4
\left(\ln\frac{\mathcal{M}^2}{\mu_r^2}-\frac{3}{2}\right)\right],
\end{align}
where $\mathcal{M}^2$ are the field-dependent mass matrices, $\mu_r$ is the renormalization scale, and the supertrace includes a factor of $(-1)^{2J}(2J+1)$ for each particle of spin J in the loop. We have omitted additional charge and colour factors which should be appropriately included. The corresponding one-loop contribution to the neutral Higgs mass matrix  is given by Eq.~(\ref{1Lmh})
\begin{align}
(\Delta\mathcal{M}^2_h)_{ij}
&=\left.\frac{\partial^2 V_{\rm{CW}}(\Phi)}{\partial \Phi_i\partial \Phi_j}\right|_{\rm{vev}}
-\frac{\delta_{ij}}{\langle \Phi_i\rangle}\left.\frac{\partial V_{\rm{CW}}(\Phi)}{\partial \Phi_i}\right|_{\rm{vev}}
\nn\\
&=\sum\limits_{k}\frac{1}{32\pi^2}
\frac{\partial m^2_k}{\partial \Phi_i}
\frac{\partial m^2_k}{\partial \Phi_j}
\left.\ln\frac{m_k^2}{\mu_r^2}\right|_{\rm{vev}}
+\sum\limits_{k}\frac{1}{32\pi^2}
m^2_k\frac{\partial^2 m^2_k}{\partial \Phi_i\partial \Phi_j}
\left.\left(\ln\frac{m_k^2}{\mu_r^2}-1\right)\right|_{\rm{vev}}
\nonumber\\
&\quad-\sum\limits_{k}\frac{1}{32\pi^2}m^2_k
\frac{\delta_{ij}}{\langle \Phi_i\rangle}
\frac{\partial m^2_k}{\partial \Phi_i}
\left.\left(\ln\frac{m_k^2}{\mu_r^2}-1\right)\right|_{\rm{vev}}\ ,\quad \Phi_{i,j}=H^0_{u,r},H^0_{d,r},S_r,T^0_r\ .
\label{1Lmh}
\end{align}
The mass of the lightest Higgs boson gets additional contributions, respect to the MSSM case, both at tree level and at one-loop. At tree level such contributions are due to the triplet and singlet scalars, as reported in Eq.~\ref{hbnd} and showed in Figure~\ref{mht}. At one-loop this is due to the presence of additional supersymmetric particles running in the loops, and this reduces the amount of corrections needed from the squarks. To illustrate this point, in Figure~\ref{mh1lvsp} we have plotted the lightest CP-even neutral Higgs mass at one-loop versus the lighter stop mass ($m_{\tilde{t}_1}$). We have used the following color coding convention: the red points are mostly doublets ($\geq 90\%$), the green points are mostly triplet/singlet($\geq 90\%$) and the blue points are mixed ones.  The yellow band shows the Higgs mass range $123\leq m_{h_1} \leq 127$ GeV. We notice that a $\sim 125$ GeV CP-even neutral Higgs could be obtained by requiring a stop of very low mass, as low as 100 GeV. Thus, in the case of extended SUSY scenarios like the TNMSSM, 
the discovery of a $\sim 125$ GeV Higgs boson does not put a stringent lower bound on the required SUSY mass scale, and one needs to rely on direct SUSY searches for that.

\section{Perturbativity limits}\label{pblmt}
\begin{figure}[bht]
\begin{center}
\mbox{\subfigure[]{\hskip -15 pt
\includegraphics[width=0.6\linewidth]{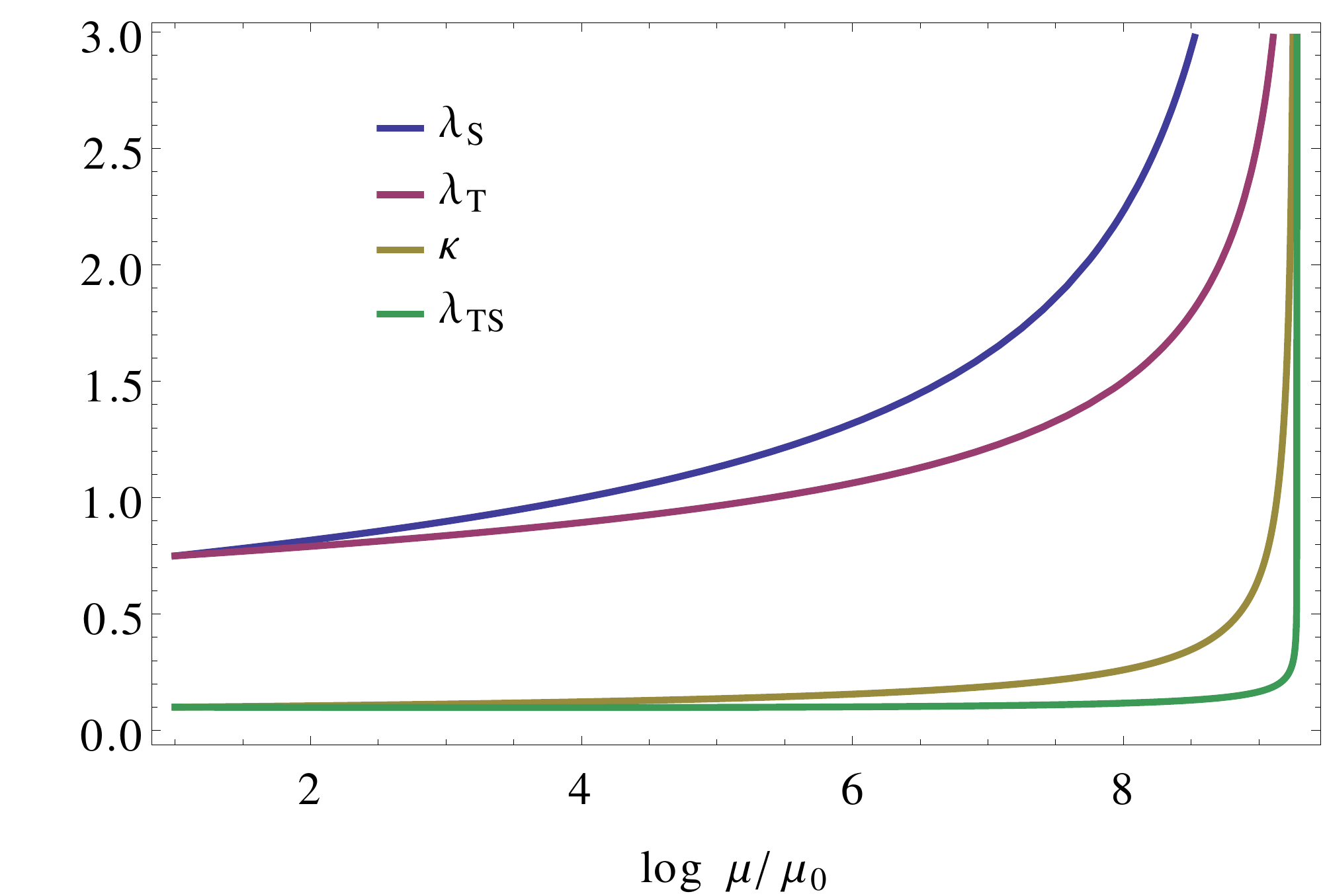}}}
\caption{The dashed lines show the evolution of the dimensionless couplings at two-loop, whereas the solid lines show the behavior at one-loop.}\label{betaf}
\end{center}
\end{figure}

The $\beta$ functions for a generic $\mathcal{N}=1$ supersymmetric theory are well known in the literature and can be obtained straightforwardly for our model \cite{sarah}. We focus our attention on the dimensionless coupling of the scalar sector, because the triplet spoils the gauge coupling unification under the renormalization group evolution.
As we show in Figure \ref{betaf}, choosing a relatively lower values of $\lambda_{TS}$, $\kappa$ and $\tan{\beta}=3$ would allow the theory to stay perturbative until $10^{8-10}$ GeV even with
$\lambda_{T, S}$ as large as $0.8$. However, the choice of a relatively large value of $\tan{\beta}$, i.e. around 10,  would extend the perturbativity scale up to $10^{15}$ GeV. The reason to have larger values of $\lambda_{T, S}$ is to increase the tree-level contributions to the Higgs mass (see Eq.~(\ref{hbnd})) as well as the radiative corrections via the additional Higgs bosons exchanged in the loops. Both of these contributions reduce the amount of supersymmetric fine-tuning, assuming a Higgs boson of $\sim 125$ GeV in the spectrum,  by a large amount, respect both to a normal and to a constrained MSSM scenario.

The minimisation conditions for the scalar potential relate the $Z$ boson mass to
the soft breaking parameters in the form  
\bea
M_Z^2&=&\mu^2_{\rm{soft}}-\mu_{\text{eff}}^2\\
\mu_{\text{eff}}&=&v_S \lambda_S-\frac{1}{\sqrt 2}v_T\lambda_T, \quad \mu^2_{\rm{soft}}=2\frac{m_{H_d}^2-\tan^2\beta\, m_{H_u}^2}{\tan^2\beta -1}.
\eea
It is also convenient to introduce the additional parameter
\bea\label{ft}
\mathcal{F}&=&\left|\ln\frac{\mu^2_{\rm{soft}}-\mu_{\text{eff}}^2}{\mu^2_{\text{soft}}}\right|,
\eea
characterizing the ratio between $M^2_Z$ and $\mu^2_{\text{soft}}$, which can be considered a measure of the fine-tuning. 
\begin{figure}[bht]
\begin{center}
\mbox{\subfigure[]{\hskip -15 pt
\includegraphics[width=0.55\linewidth]{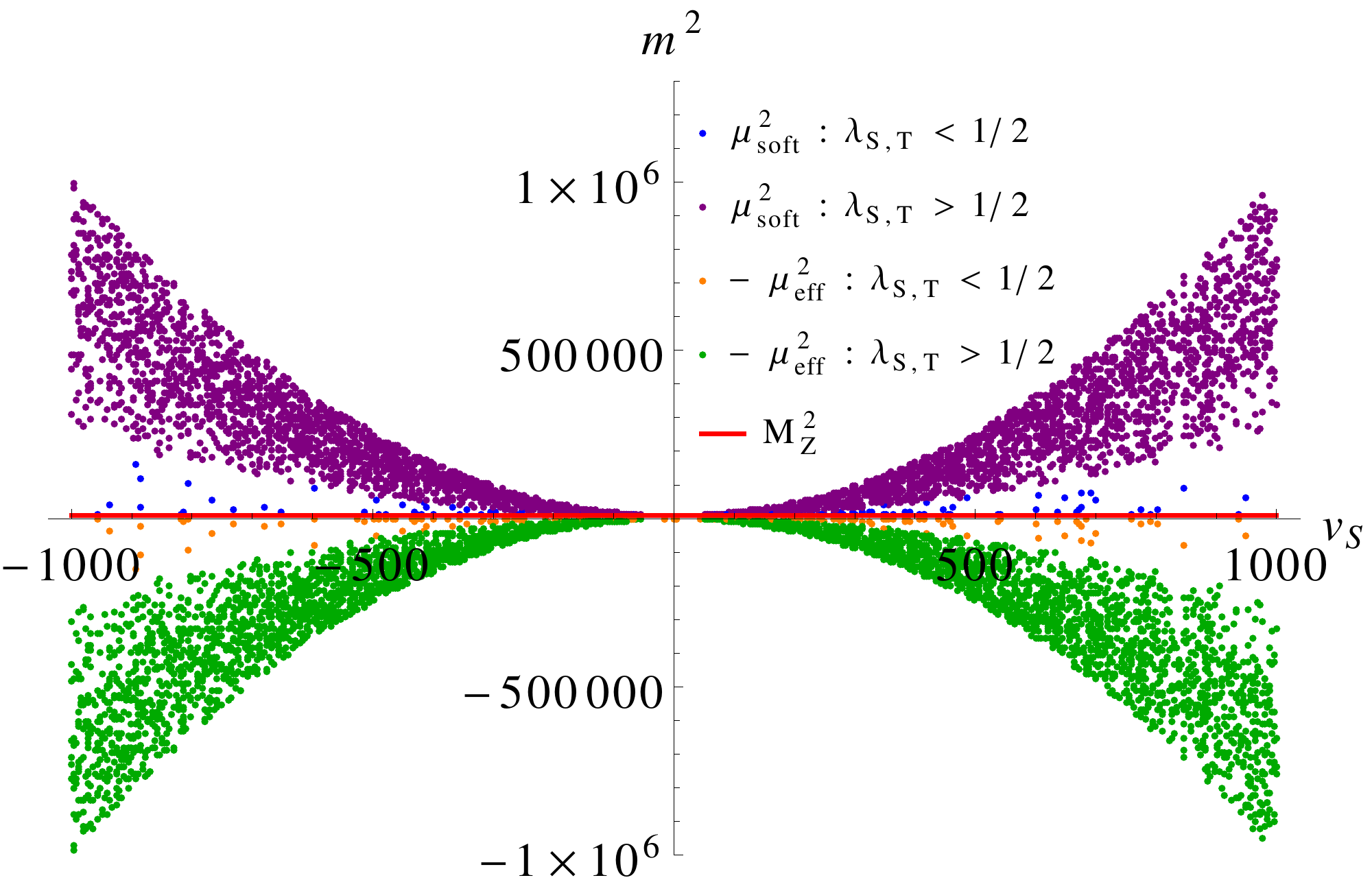}}
\subfigure[]{\includegraphics[width=0.55\linewidth]{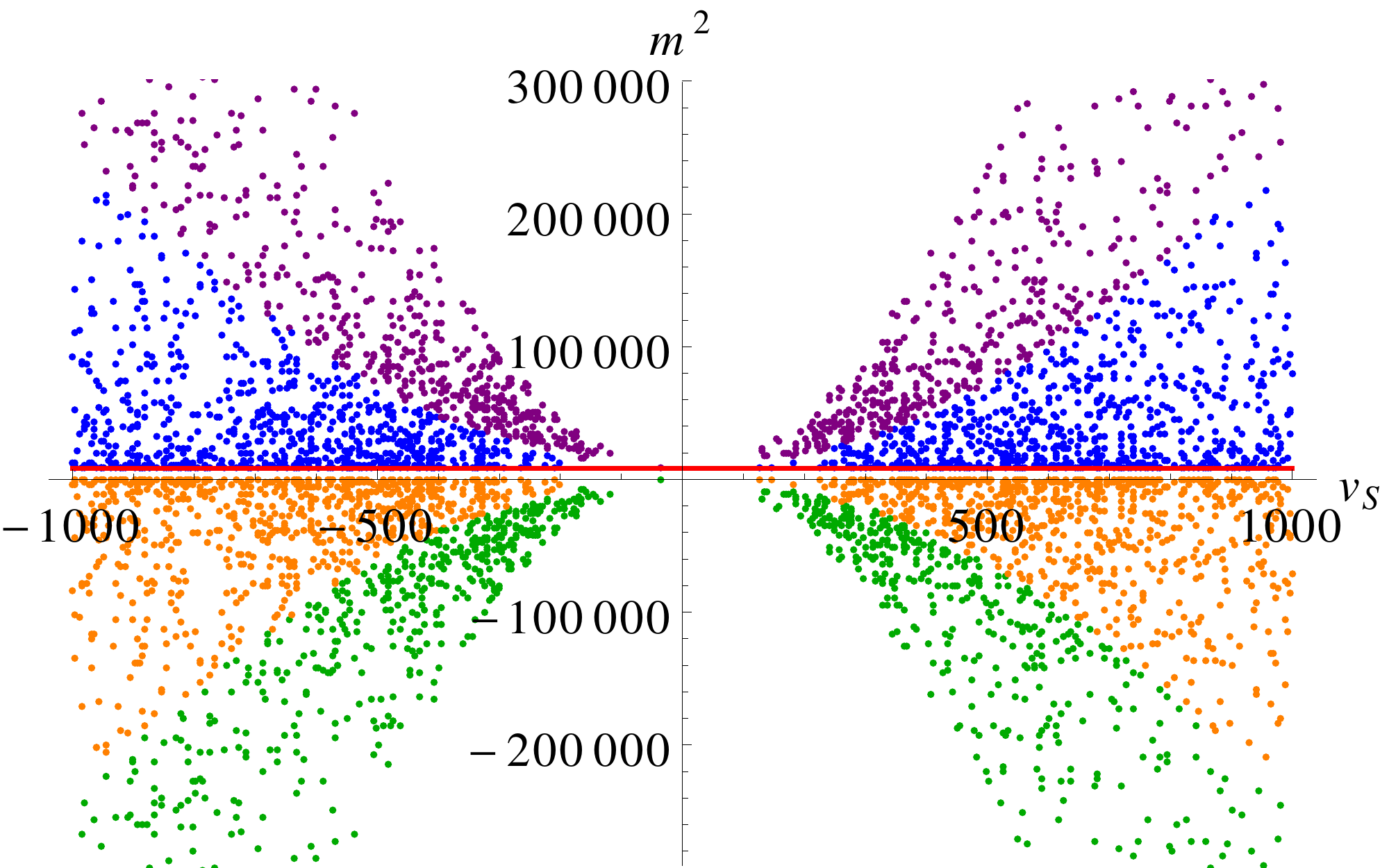}}}
\caption{The (a)tree-level and (b) one-loop level fine-tuning measures $\mu_{\text{soft}}$ and  $-\mu^2_{\text{eff}}$ versus the singlet vev $v_S$ for a candidate Higgs of mass between $120\leq m_{h_1} \leq 130$ GeV respectively. The violet points represent $\mu^2_{\rm{soft}}$ values $\lambda_{S,T}\geq 0.5$ and the blue points represent $\mu^2_{\rm{soft}}$ values $\lambda_{S,T}<0.5$. The green points represent $\mu^2_{\rm{eff}}$ values with $\lambda_{S,T}\geq 0.5$ and the orange points indicate $\mu^2_{\rm{eff}}$ values $\lambda_{S,T}< 0.5$. The red line shows the $Z$ boson mass $M_Z$ \cite{TNSSMo}.}\label{fnt}
\end{center}
\end{figure}

As reported in \cite{TNSSMo},  Figure~\ref{fnt} shows $\mu^2_{\text{soft}}$ and  $-\mu^2_{\text{eff}}$ versus the singlet vev $v_S$ for tree-level candidate Higgs masses in the interval $120\leq m_{h_1} \leq 130$ GeV, as well as at one-loop. The violet points represent $\mu^2_{\rm{soft}}$ values for which $\lambda_{S,T}\geq 0.5$, and the points in blue refer to values of $\mu^2_{\rm{soft}}$ with $\lambda_{S,T}<0.5$. The green points mark values of $\mu^2_{\rm{eff}}$ with $\lambda_{S,T}\geq 0.5$, and the orange points refer to $\mu^2_{\rm{eff}}$ values with $\lambda_{S,T}< 0.5$. The high $\lambda_{T,S}$ points are much needed for the tree-level result to attain  the Higgs mass around 125 GeV, and are significantly fine-tuned ($\mathcal{F}\sim 5$) respect to the $Z$ mass.  The low $\lambda_{T,S}$ points
correspond to less fine-tuning, but  are less numerous at tree-level, ($\mathcal{F}\lesssim 2$). Their number increases at one-loop order, due to the extra radiative contributions to the Higgs mass.

\section{Possibility of a light pseudoscalar}\label{lpss}

In the limit when the parameters $A_i$ in Eq.~(\ref{softp}) go to zero, the  discrete $Z_3$ symmetry of the Lagrangean is promoted to a continuos $U(1)$ symmetry given by Eq.~(\ref{csmy}) 
\bea\label{csmy}
(\hat{H}_u,\hat{H}_d, \hat{T},\hat{ S}) \to e^{i\phi}(\hat{H}_u,\hat{H}_d, \hat{T},\hat{ S}) .
\eea 
An explicit breaking of a continuous global symmetry is expected to be accompanied by pseudo-Nambu Goldstone bosons (pNGB) as in the case of chiral symmetry breaking in QCD where the pions take the role of the corresponding pNGB. In this case, in the TNMSSM we should expect a light pseudoscalar in the spectrum, whose mass is of the same order of the $A_i$ parameters. A similar behaviour is allowed in the  NMSSM and the corresponding light pseudoscalar is known as the $R$-axion \cite{nmssm, Agashe:2011ia}.  The mass of such a very light pseudoscalar does not receive any direct bound, as long as it satisfies the minimization conditions. However, indirect bounds emerge if this light pseudoscalar decays into fermions.  Previous analysis at LEP have searched for such light scalar bosons in
$e^+ e^-\to Z h$ and $e^+ e^-\to A h$, where $h, A$ are CP-even and odd neutral Higgs bosons respectively \cite{LEPb}. Similar bounds also come from the bottomonium decay \cite{bottomium} for such light pseudoscalar in the mass range of 5.5 -14 GeV. Recently published data from CMS also provide strong bounds for such a light pseudoscalar when it couples to fermions \cite{cmsab}. 

\begin{figure}[h]
\begin{center}
\mbox{\subfigure[]{\hskip -15 pt
\includegraphics[width=0.55\linewidth]{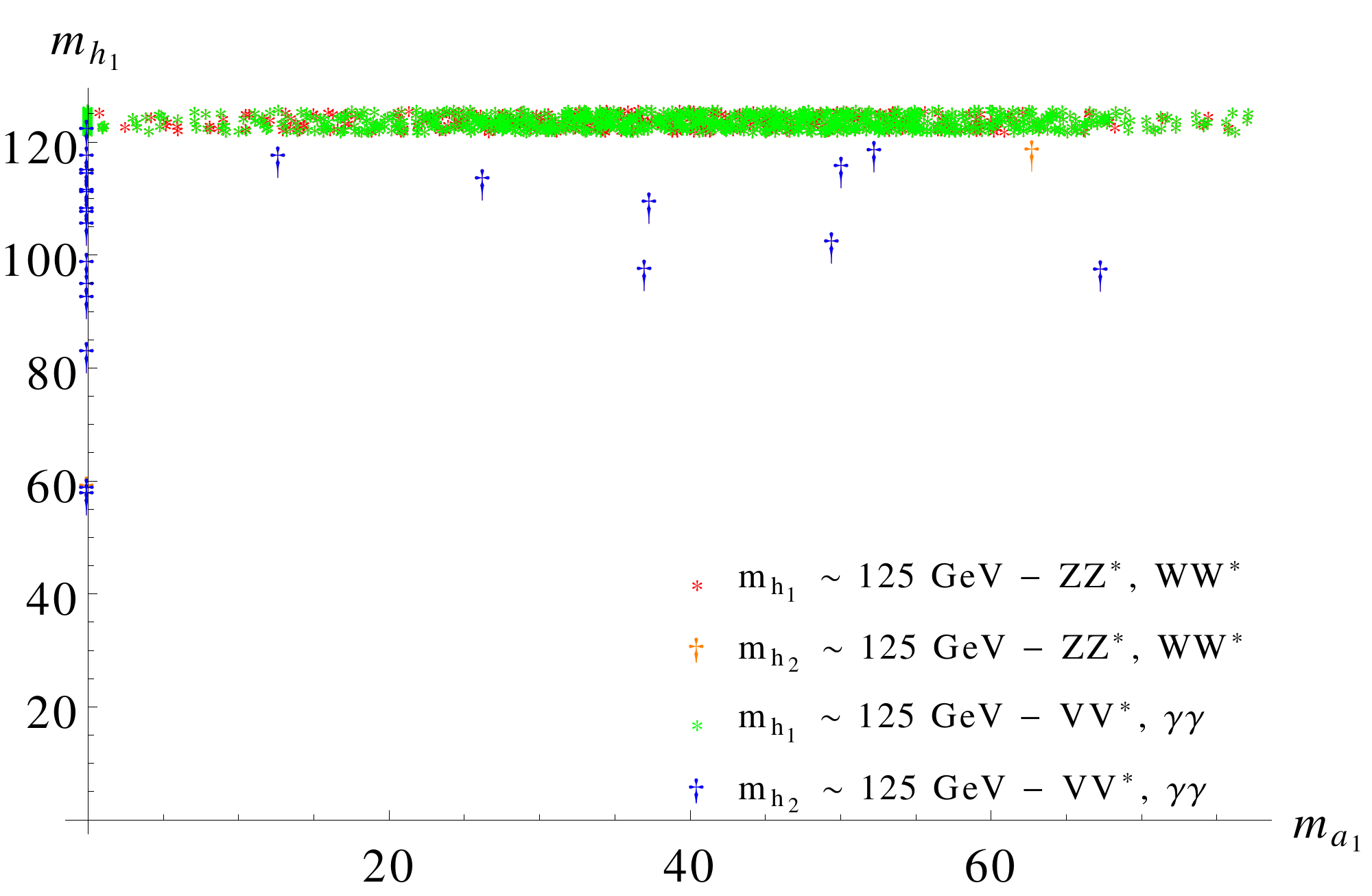}}
\subfigure[]{\includegraphics[width=0.55\linewidth]{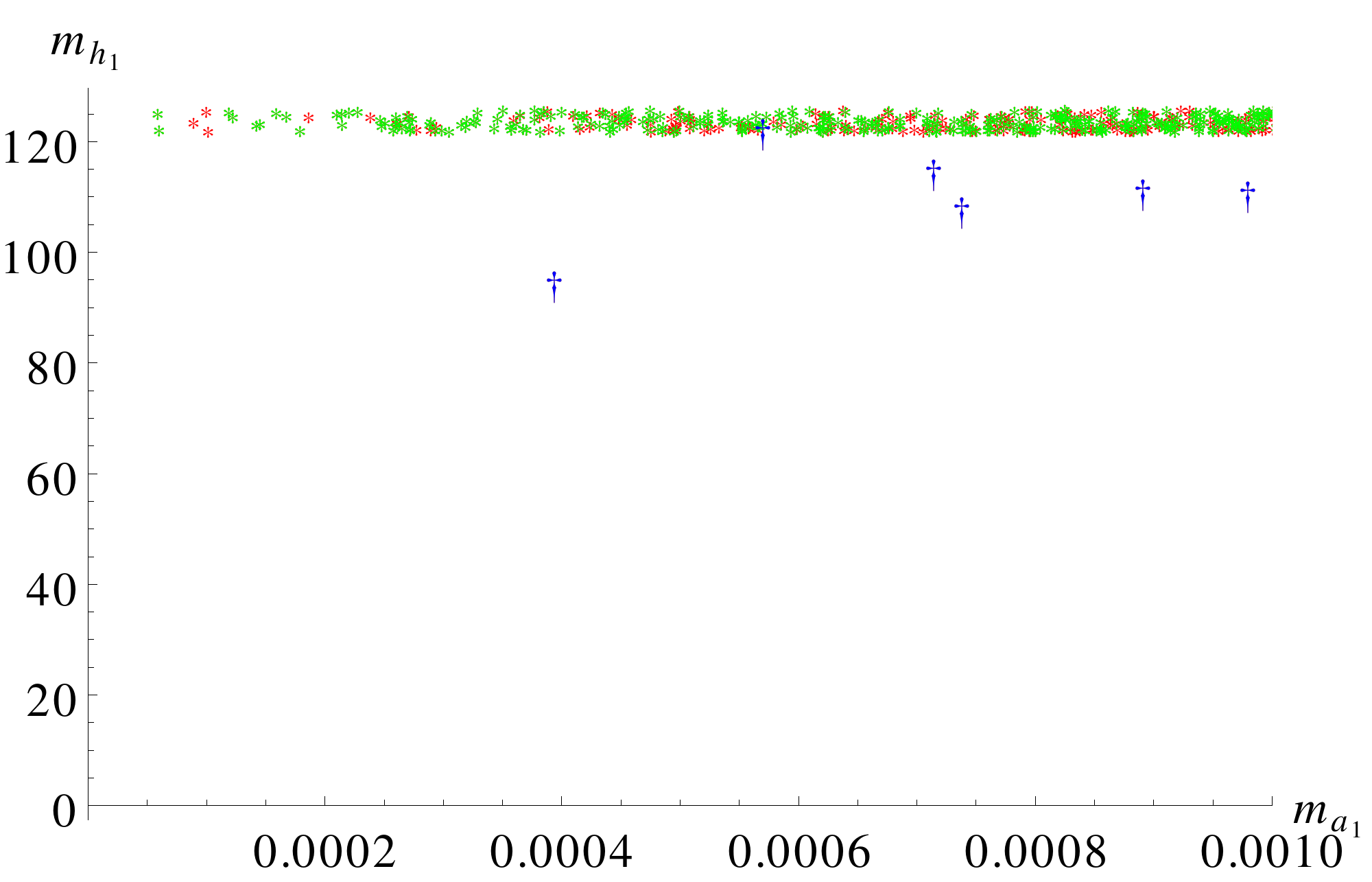}}}
\caption{The lightest CP-even Higgs boson mass $m_{h_1}$ vs the lightest pseudoscalar mass $m_{a_1}$ at one-loop (top-stop and bottom-bottom corrections) consistent with the Higgs data from CMS, ATLAS and LEP. The red points corresponds to the case where $m_{h_1}\sim m_{125}$, the orange points correspond to mass values of $m_{h_1}$ and $m_{a_1}$  where  $m_{h_2}\sim m_{125}$ and all of them satisfy  the $ZZ^*$, $WW^*$ bounds at $1\sigma$ and $\gamma\gamma$
bound at $2\sigma$ level from both CMS and ATLAS. The red (orange) points which satisfy the  $\gamma\gamma$ result at $1\sigma$ are marked green (blue). Very light pseudoscalar masses $m_{a_1}\leq 1$ MeV are shown in panel (b), which is a zoom of the small mass region of (a) \cite{TNSSMo}.}\label{higgsdata}
\end{center}
\end{figure}

If the light pseudoscalar in the TNMSSM is of singlet/triplet-type then it doesn't couple to the fermions and to the $Z$ boson, which makes the corresponding (green) points to be allowed by the data. For this purpose we have calculated the neutral Higgs boson mass spectrum at one-loop via the Coleman-Weinberg prescription for the effective potential \cite{TNSSMo}. After
computing the Higgs boson mass spectrum, we check if a given parameter point allows a CP-even Higgs boson around $125$ GeV.  The points considered are those with a Higgs $\sim 125$ GeV which at the same time satisfy the $ZZ^*$, $WW^*$ bounds at $1\sigma$ level and the $\gamma\gamma$ bound at $2\sigma$ level from both CMS and ATLAS. They are marked in red (orange) in Figure~\ref{higgsdata}. 
The red (orange) points which satisfy the  $\gamma\gamma$ result at $1\sigma$ are marked in green (blue).  The allowed mass values are shown as red points for which the lightest CP-even Higgs boson ($h_1$) is the detected Higgs at $\sim 125$ GeV. In this case there is one hidden Higgs boson, i.e. a light pseudoscalar with a mass less than 125 GeV. The orange points correspond to a scenario where $m_{h_2}\sim m_{125}$ and which leaves both  $h_1$ and $a_1$ hidden ($< 125$ GeV). These requirements automatically bring the fermionic decay modes closer to the SM expectation. Of course the uncertainties of these decay widths leave a room for $h_{125}\to a_1a_1/h_1h_1$ decays. 

 For much smaller values of $A_i \sim 0$, the pNGB gets a very small mass. Figure~\ref{higgsdata}(b) 
 shows  such solutions where $m_{a_1} \leq 1$ MeV. The points in this case correspond to possible $a_1$ states which do not decay into any charged fermion pair ($m_{a_1} \leq 2m_{e}$) and have an interesting phenomenology. Such light pseudoscalars are then only ones allowed to decay into two photons via doublet mixing mediated by a fermion loop. If its lifetime is greater than the age of the universe, then the $a_1$ is a possible dark matter candidate. Two hidden Higgs bosons render the phenomenology very interesting, allowing both the $h_{125} \to a_1 a_1$ and the $h_{125} \to h_1 h_1$ decay channels discussed in \cite{TNSSMo,Bandyopadhyay:2015tva}. 

\section{ LHC Phenomenology}\label{pheno}
\begin{figure}[hbt]
\begin{center}
\includegraphics[width=0.6\linewidth]{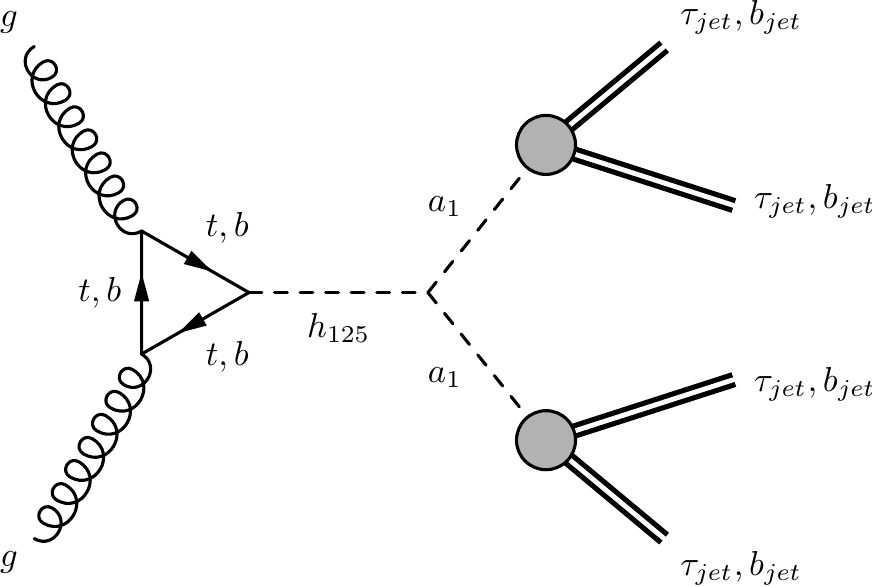}
\caption{Pseudoscalar (triplet/singlet) pair production from Higgs boson, generated via 
gluon-gluon fusion, and their decays via their mixing with the doublets \cite{Bandyopadhyay:2015tva}.}\label{glutri}
\end{center}
\end{figure}

In the previous section we have shown that it is possible to have a light pseudoscalar ($\leq 100$ GeV) which
is mostly singlet/triplet-like. Being mostly singlet-like, it is difficult to produce it directly at a hadron collider. It couples to the doublet-like Higgs bosons via $\lambda_S$ (see Eq.~\ref{spt}). This makes it easy to produce by an intermediate Higgs. If such a light pseudoscalar ($m_{a_1}\le 125/2$) exists, then it can be produced in $gg \to h_{125} \to a_1 a_1$ as shown in Figure~\ref{glutri}.  Such pair production can give rise to final states rich in $\tau$, $b$ and even in muons. In \cite{Bandyopadhyay:2015tva} we have investigated such final states by considering all the dominant SM backgrounds at the LHC with a center of mass energy of 13 and 14 TeV. A detailed signal to background analysis of the final states $2\tau + 2b$, $2b +2\mu$ and  $\geq 3\tau$ reveals that some of the benchmark points with the light pseudoscalar ($m_{a_1}\sim 10-20$ GeV) can be probed with early data of 25 fb$^{-1}$ at the LHC.

\begin{figure}[bht]
\begin{center}
\includegraphics[width=0.33\linewidth, angle=-90]{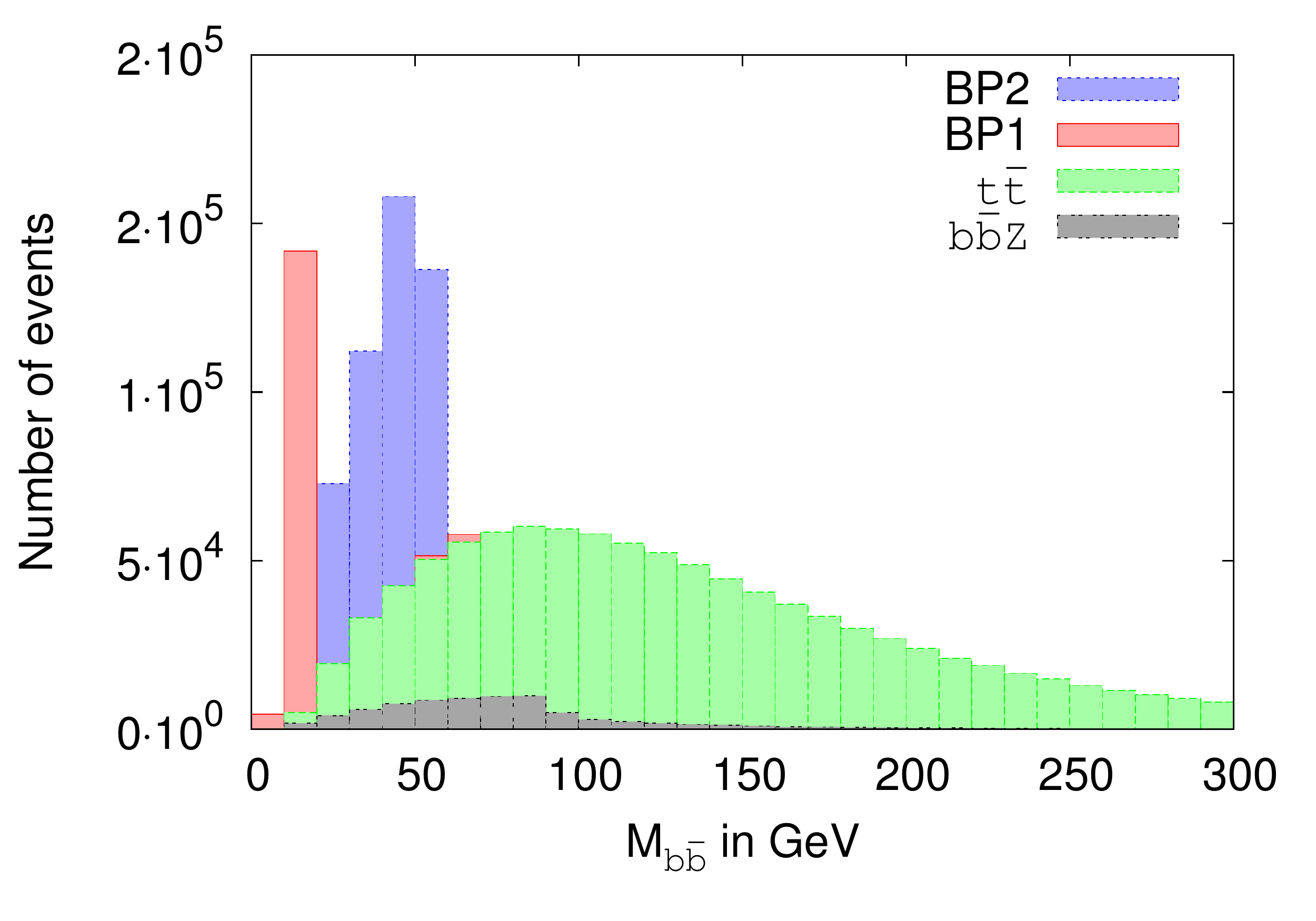}
\includegraphics[width=0.33\linewidth, angle=-90]{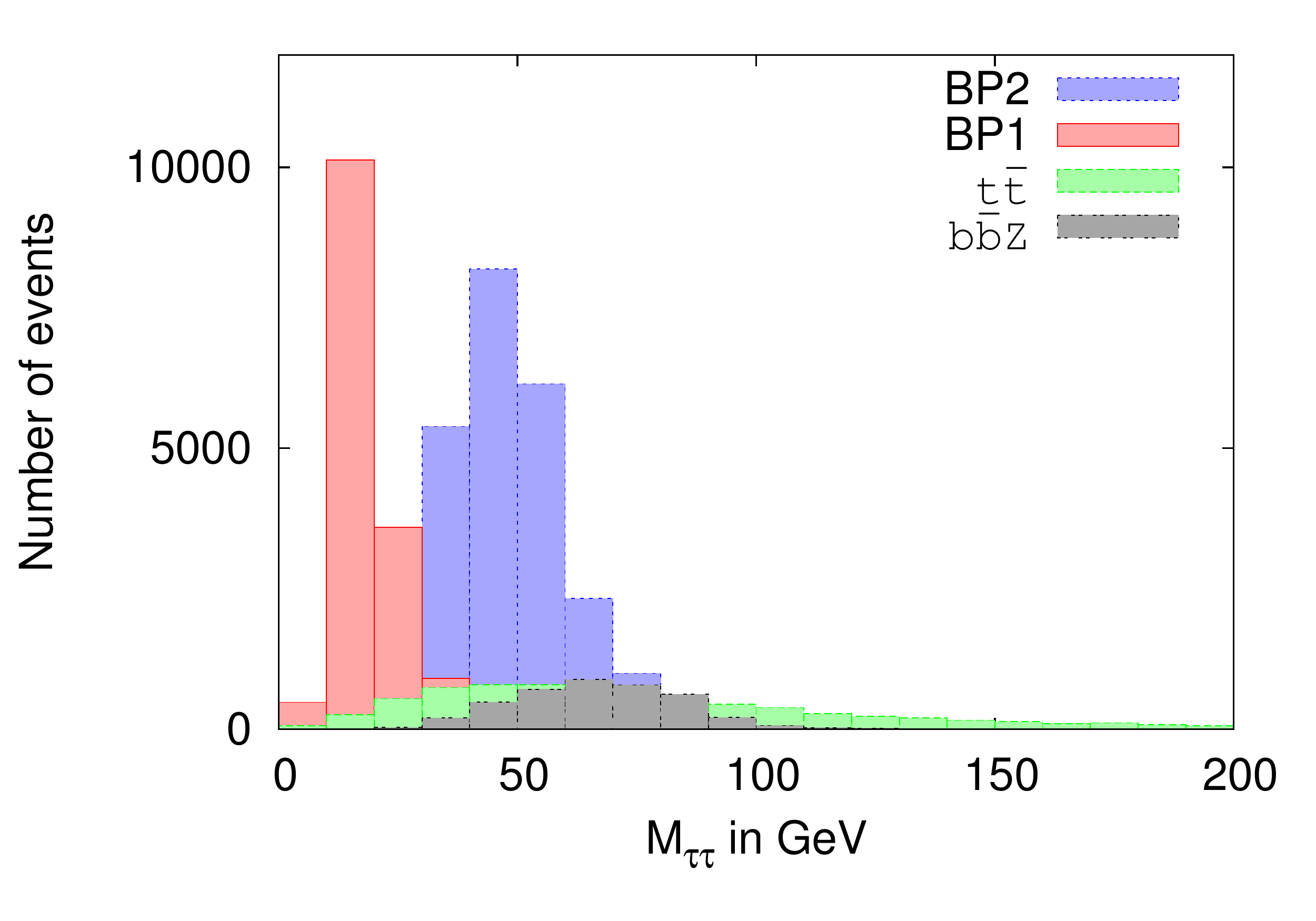}
\caption{ Invariant mass distribution of $b_{\rm{jet}}$'s (left) and $\tau_{\rm{jet}}$'s (right) for $t\bar{t}$ and for the benchmark points discussed in \cite{Bandyopadhyay:2015tva}.}\label{invdis}. 
\end{center}
\end{figure}

The presence of a light pseudoscalar can also be probed from the charged Higgs decay, i.e. $h^\pm_1 \to a_1 W^\pm$. The light pseudoscalar is also present in the case of the NMSSM with a $Z_3$ symmetry. The difference between the TNMSSM and the NMSSM in the charged Higgs sector originates as the former has three charged Higgs bosons, among which two could be of triplet type. It has been shown that the non-standard decay mode of the light charged Higgs boson $h^\pm \to a \,W^\pm$ can evade the recent mass bounds from LHC \cite{chLHC} and can be probed with the $\tau$ and $b$ rich final states coming from the light pseudoscalar \cite{chaW}. The charged Higgs boson in this case (NMSSM) is still doublet-like so its production from the decay of a top or via $b g\to t h^\pm$ is still possible. The possibility of a light scalar 
is also present in the MSSM with CP-violating interactions, where the light Higgs is mostly CP-odd and can evade the existing LEP bounds. Such Higgs boson with a mass $\lesssim 30$ GeV can open the decay mode $h^\pm \to h_1 W^\pm$,  which can be explored at the LHC\cite{cpv}.\\
However things can change a lot in a triplet extended supersymmetric model (TESSM), where we have a light charged Higgs triplet. It does not couple to fermions, which makes its production via conventional methods a lot harder. However, its production channels in association with $W^\pm$ bosons or with neutral Higgs bosons are still considerable. The interesting feature of having a charged Higgs triplet is the non-zero $h^\pm_i - Z- W^\mp$ coupling, which not only gives rise to a new decay mode of $ZW^\pm$ but also makes the charged Higgs production possible via vector boson fusion \cite{tripch}. This is not allowed for the doublet-like charged Higgs boson of the MSSM or of the 2HDM. In the TNMSSM we have both the presence of a very light pseudoscalar - as in the NMSSM -  and of a charged Higgs triplet, as in the TESSM. If we can pair produce such charged Higgs where one of them decays into $a_1 W^\pm$ and the other into $ZW^\pm$, then we can search for such decay modes. This provides a smoking gun
signature for the extended Higgs structure of the TNMSSM.  We are currently investigating in detail ways to probe both aspects at the LHC \cite{Chsim}.
\section{Conclusions}\label{concl}
The discovery of a Higgs boson around 125 GeV in mass has opened the debate if extra Higgs bosons are present in Nature. They may belong to higher representations of the same $SU(2)$ gauge symmetry or of a different one. 
In particular, the question of the existence of a very light scalar below 100 GeV is still a theoretical possibility which can be addressed in the context of various representations. Given that all the SM decay modes of the Higgs boson around $125$ GeV are yet to be discovered, the question of the existence of non-standard decays in this sector still remains elusive. The other Higgs bosons could have higher mass values  ($> 125$ GeV) but the possibilities of lower mass values ($< 100$ GeV) also remain quite open. Certainly in such cases, if the $h_{125}\to h_1h_1/a_1a_1$ decay modes are 
kinematically allowed, then they can be probed at the LHC. \\
Models with $Z_3$ symmetry, such as the NMSSM or the TNMSSM, where such scalars are naturally light, can be tested once additional data will be made available from the LHC. If an extended Higgs sector exists, then finding a charged Higgs is a direct proof of it. So far, the search for a charged Higgs boson is performed in its decay to $\tau\nu$, but the existence of a light scalar/pseudoscalar gives us the new decay mode $h^\pm \to a_1/h_1 W^\pm$, which should be looked for at the LHC. Triplet charged and neutral Higgs bosons do not couple to fermions and a charged Higgs triplets decay to $ZW^\pm$. These are some of the new features which can be explored at the LHC and at future colliders. To complete our understanding of electroweak phase transition we need to explore all the theoretical possibilities experimentally, with the exclusion of none. Most certainly, hidden or buried Higgs bosons can be probed at the LHC@14 TeV with additional data. The 750 GeV resonance in the diphoton mode has surely raised new expectations, for being a new physics signal. 
The answer to whether this signal can be explained by an extra Higgs or not requires more data and the exploration of other decay channels.

\section{Acknowledgement} 
PB wants to thank CORFU 2015 organiser for the Young Scientist award and INFN Lecce for
partial support to attend the workshop. The work of C.C. is supported by a {\em The Leverhulme Trust Visiting Professorship} at the STAG research Center and Math Sciences of the University of Southampton.

\end{document}